\documentclass[preprint,12pt]{elsarticle}

\usepackage{amssymb}
\usepackage{amsmath}

\usepackage{amsmath,amsfonts}

\usepackage{algorithmic}
\usepackage{algorithm}
\usepackage{array}
\usepackage{subcaption}
\usepackage{rotating}

\usepackage{textcomp}
\usepackage{stfloats}
\usepackage{url}
\usepackage{verbatim}
\usepackage{graphicx}
\usepackage{stmaryrd}
\usepackage{comment}
\usepackage{amssymb, amsmath, bm}
\usepackage{mathrsfs}
\usepackage[dvipsnames]{xcolor}
\usepackage{multirow}
\usepackage{booktabs, tabularx}

\definecolor{myred}{RGB}{162, 0, 37}

\journal{Computer Speech \& Language}

\begin{document}

\begin{frontmatter}



\title{Quantized Approximate Signal Processing (QASP):
Towards Homomorphic Encryption for audio}


\author[callyope]{Tu Duyen Nguyen} 
\author[callyope]{Adrien Lesage}
\author[callyope]{Clotilde Cantini}
\author[callyope]{Rachid Riad}
\affiliation[callyope]{organization={Callyope},
            city={Paris},
            country={France}}

\begin{abstract}
Audio and speech data are increasingly used in machine learning applications such as speech recognition, speaker identification, and mental health monitoring. However, the passive collection of this data by audio-listening devices raises significant privacy concerns. Fully homomorphic encryption (FHE) offers a promising solution by enabling computations on encrypted data and preserving user privacy. Despite its potential, prior attempts to apply FHE to audio processing have faced challenges, particularly in securely computing time-frequency representations—a critical step in many audio tasks.

Here, we addressed this gap by introducing a fully secure pipeline that computes, with FHE and quantized neural network operations, four fundamental time-frequency representations: Short-Time Fourier Transform (STFT), Mel filterbanks, Mel-frequency cepstral coefficients (MFCCs), and gammatone filters. Our methods also support the private computation of audio descriptors and convolutional neural network (CNN) classifiers. Besides, we proposed approximate STFT algorithms that lighten computation and bit use for statistical and machine learning analyses. We studied these additional approximations theoretically and empirically.

We ran experiments on the VocalSet and OxVoc datasets demonstrating the fully private computation of our approach. We showed significant performance improvements with STFT approximation in private statistical analysis of audio markers, and for vocal exercise classification with CNNs.  Our results reveal that our approximations substantially reduce error rates compared to conventional STFT implementations in FHE. We also demonstrated a fully private classification based on the raw audio for gender and vocal exercise classification. Finally, we provided a practical heuristic for parameter selection, making quantized approximate signal processing accessible to researchers and practitioners aiming to protect sensitive audio data.

\end{abstract}


\begin{highlights}

\item First complete end-to-end demonstration of fully homomorphic encryption (FHE) applied directly to raw audio signals, ensuring complete privacy during inference. 
\item Novel quantized approximate signal processing techniques for secure computation of STFT, Mel filterbanks, MFCCs, and gammatone filters.
\item Theoretical error bounds established for approximation methods, guiding optimal parameter selection.
\item Empirical validation on multiple tasks (spectrogram reconstruction, vocal marker analysis, and machine learning classification) and multiple datasets demonstrates effective private audio processing with minimal performance loss.
\end{highlights}


\end{frontmatter}


\section{Introduction}
 The number of audio-listening devices has surged thanks to the increasing affordability of smart speakers, headphones, and even TVs, putting high-fidelity audio capture technology in the hands of consumers at ever-lower costs. This trend, while allowing easy access to smart agents through audio channels, has consequences for user privacy. 

Human speech signal conveys sensitive information beyond linguistic content about the speaker's traits and current state \citet{narayanan2013behavioral}. Automatic speech systems have been developed to recognize personal traits such as age, gender \citet{hechmi2021voxceleb}, height \citet{mporas2009estimation}, current emotions \citet{schuller2018speech, akccay2020speech}, mood states in psychiatric diseases \citet{gideon2016mood, cummins2015review}, or even current pain \citet{ren2018evaluation}. The increasing prevalence of devices equipped with microphones expands the possibilities for adversaries to capture speaker information. This proliferation of devices can be seen through the lens of the cryptography principle known as the "surface attack" theory \citet{howard2005measuring}. A larger attack surface – the total number of potential entry points for adversaries – implies a larger number of vulnerabilities.

This ever-growing attack surface on human speech calls for the development and deployment of machine learning and speech technologies to preserve the privacy of speakers \citet{nautsch2019preserving}. Given the sensitivity of speech data, individuals may wish to protect both their voice identity and the content of their utterances. Such privacy concerns are often reinforced by legal frameworks like the EU's GDPR, which mandate the protection of personal data. This is even more critical in healthcare settings, where speech analysis is gaining traction in neurology and psychiatry \citet{cummins2015review, riad2022predicting, fraser2016linguistic}, often through applications developed by private companies. 
Privacy-preserving techniques must be implemented throughout the entire machine learning pipeline to ensure full protection of individuals' speech. This includes protecting speech data during training data collection, model inference (prediction), and even potential privacy breaches by cloud vendors and healthcare companies hosting and carrying machine learning analyses (see \citet{backstrom2023privacy,nautsch2019preserving} for reviews about vulnerabilities, privacy-preserving methods for speech).

While existing methods like differential privacy \citet{pelikan2023federated}, speech anonymization \citet{srivastava2022privacy}, and federated learning \citet{kairouz2021advances} aim to protect training speaker data, they have some limitations and do not protect data used during deployment and inference. These techniques will decrease potential leakages of the training data, by reducing speaker footprints \citet{shamsabadi2023differentially} on spoken utterances; or leaving training data on mobile smartphones \citet{guliani2021training}, and computing some gradients on the client side. 

These approaches present some limitations. First, they can conflict with biometric or clinical speech applications. For example, deleting speaker characteristics like pitch or speech rate can hinder tasks like emotion recognition or disease severity estimation. Second, these methods still have security risks. Storing model weights on mobile devices exposes training data participants to membership inference attacks \citet{teixeira2024improving} and even potential data reconstruction \citet{rigaki2023survey}. 

A solution to circumvent such risks is to encrypt the sensitive audio and use it to perform the computations with homomorphic encryption.
The current demonstrations and applications of homomorphic encryption are restricted to linear models, tree-based models \citet{frery2023privacy} or shallow neural networks \citet{stoian2023deep}. This is limited by integer-based representation of numbers and memory limits to perform computation in homomorphic domain. To circumvent the problems of a large number of computations due to signal processing, \citet{zhang2019encrypted} proposed to avoid filterbanks by not computing them and using only smaller convolutions on audio. Yet, signal processing algorithms still exhibit strong performances: Mel filterbanks are still used in state-of-the-art large models tackling automatic speech recognition like Whisper \citet{radford2023robust}. \citet{dumpala2021sine}, also explored speaker identity masking using sine-wave speech for depression detection, but there are no guarantees of the loss of information for all applications.  
A compromise in terms of security is to allow computation of time-frequency representations in clear, and a part of client pre-processing on-device and encrypting only subsequent algorithms:\citet{glackin2017privacy} adopted this strategy, and used a convolutional neural network based on the short-time Fourier transform to compute locally phonetic probabilities and encrypted the rest of the computations. 

Pioneering efforts in cryptographic-based secure speech processing were undertaken by \citet{pathak2011privacy, pathak2012privacy}. They developed algorithms for secure Gaussian mixture model computations based on the clear computation of Mel-Frequency Cepstral Coefficients (MFCCs). \citet{thaine2019extracting} used homomorphic encryption to encrypt MFCCs and Bark-Frequency Cepstral Coefficients (BFCCs), and evaluated Automatic Speech Recognition Systems on the decrypted results.
Recently, \citet{nautsch2018homomorphic} applied homomorphic encryption to securing biometric speech processing, while Treiber et al. and Nautsch et al. also explored secure multiparty computation  \citet{nautsch2019privacy, treiber2019privacy} for the same task.These approaches still assume the computation of speech representation locally also sharing service-provider models directly to users' devices. Building upon these advances, \citet{teixeira2023privacy} introduced a method to safeguard both the speaker's x-vector template and the underlying model itself. 

Encryption of the raw audio signal before sending it to the server can unlock a broader range of privacy-preserving techniques. This approach not only protects the sensitive audio data but also safeguards the pipeline and models of service providers from unauthorized access or tampering.

In this work, we investigated and introduced novel methods to address the challenge of full privacy-preserving at the \textit{inference stage} in speech and audio processing, i.e. guarantee privacy during data fed to models during deployment. Our work is in contrast with privacy-preserving techniques concerning training data protection (e.g. federated learning, differential privacy, membership inference). The challenge of this problem is to achieve privacy guarantees for speech data during model inference, while simultaneously maintaining the performance of speech analyses. We achieved this by leveraging advances in homomorphic encryption \citet{gentry2009fully, Concrete}, neural network quantization \citet{brevitas}, and signal processing properties. Similar techniques leveraging homomorphic encryption and quantization have already shown promising results in secure image compression and processing \citet{FHE_image}. 

Our first contribution is the introduction of the first system to perform homomorphic encryption computations on the raw audio signals, enabling secure processing of audio and speech with the main algorithms Short-Time Fourier Transform (STFT), Mel filterbanks, Mel-Frequency Cepstral Coefficients (MFCCs), and gammatone filters. Second, we proposed and combined multiple techniques for computing approximate time-frequency representations with low-bit depth, optimizing audio processing while maintaining better performance for more complicated tasks. We also established theoretical error bounds for each approach as a function of approximation parameters. Finally, we conducted comprehensive evaluations, benchmarking the proposed privacy-preserving techniques against clear computation on various datasets and tasks. Our evaluation framework incorporated increasingly difficult tasks, from the absolute errors in spectral features to the estimation of audio descriptors that are statistically relevant for audio tasks, and finally fully encrypted individual machine learning predictions.

This paper starts by presenting the necessary background in homomorphic encryption, quantization and signal processing to formulate our secure audio processing goal (Section \ref{section:background}). Our approximate quantized signal processing method is introduced in Section \ref{sec:qasp}. Our proposed approximate time-frequency formulations for optimizing bit usage and their corresponding error bounds are presented in Section \ref{sub_section:dilation} to Section \ref{sub_section:cropping}. Our evaluation framework and experimental results are shown in Section \ref{section:experiments}.

\section{Background}\label{section:background}

\subsection{Audio signal processing}\label{section:context}

Deep learning models trained on large datasets are the current paradigm to perform various audio processing tasks, such as voice activity detection \citet{voiceactivitydetection}, speaker recognition \citet{Xvectors, Chung2018VoxCeleb2DS}, speech recognition \citet{radford2023robust} or emotion recognition \citet{schuller2018speech}. Even state-of-the-art models still use signal processing algorithms filterbanks \citet{radford2023robust, gong21b_interspeech} as input instead of the raw waveform. 

These algorithms usually start by mapping the raw audio signal to a representation in a time-frequency domain. They do so by reducing time resolution to gain localized information on the frequency energy. This allows the extraction of localized patterns in both time and energy which are relevant for downstream tasks. Different time-frequency representations have been proposed, heavily inspired by the human auditory system, such as Mel filterbanks, Mel-Frequency Cepstrum Coefficients (MFCCs) \citet{MFCC} or gammatone filters \citet{gammatone}. In the remainder of this work, we use interchangeably the terms \textit{time-frequency representations}, \textit{audio features} and \textit{spectrograms} to refer to these first computation steps. 

In this work, we focused on the following time-frequency representations: Short-Time Fourier Transform (STFT), Mel filterbanks, Mel-Frequency Cepstrum Coefficients (MFCCs), and Gammatone filterbanks. A secure audio processing pipeline should preserve the richness of time-frequency representations for computing tasks. To do so, we leveraged fully homomorphic encryption, quantization, sparsity and approximate signal processing. 

Time-dependent spectral analysis is usually performed by splitting the signal in successive frames and performing the discrete Fourier transform (DFT) on each frame \citet{STFT}, resulting in the short-time Fourier transform (STFT). The frame by frame concatenation of the spectral representations forms a spectrogram. A \textit{window function} is used to smooth the transition from one frame to the next.
 
Let $x \in \mathbb{R}^M$ be a real-valued signal, $h \in \mathbb{N}$ the hop length, 
$w : \mathbb{R} \rightarrow \mathbb{R}$ the Hann window with length $N$ defined by:
\begin{equation*}\label{def:stft_window}
    w(n) = \frac{1}{2}\left[1-\cos \left(\frac{2 \pi n}{N}\right)\right] 1_{0\leq n \leq N}.
\end{equation*}
We use the sliding DFT implementation of the discrete STFT: 
\begin{equation}\label{def:stft}
    X(m, k) = \sum_{n = 0}^{M-1} x(n) w(n-mh) \exp\left( \frac{-2j \pi kn}{N}\right).
\end{equation}

\subsection{Secure computation with fully homomorphic encryption}

\begin{figure}
    \centering
    \includegraphics[width=0.7\linewidth]{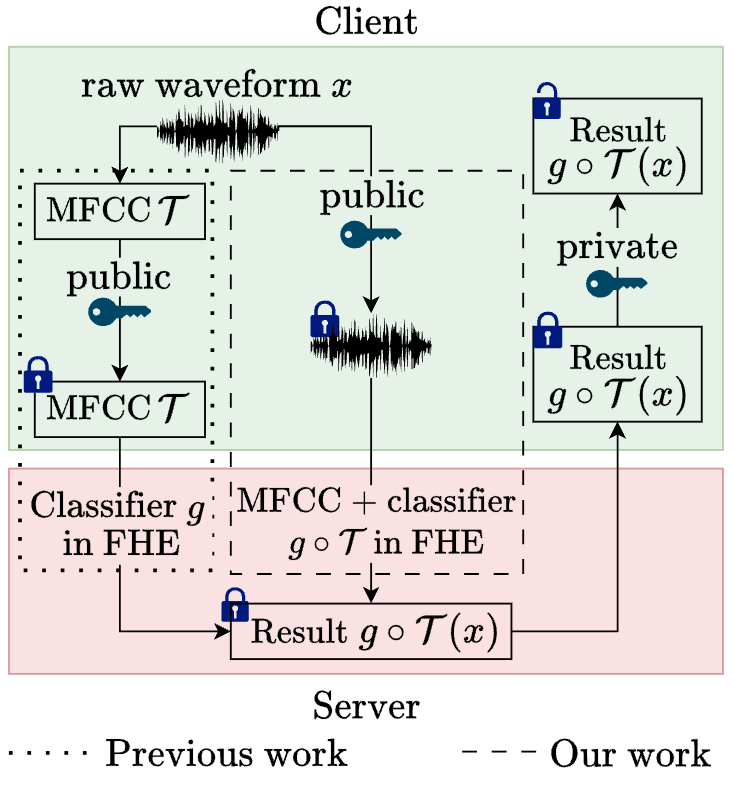}
    \caption{Schematic representation of secure audio processing with fully homomorphic encryption (FHE): client audio is not accessed in plaintext, and can be processed in encrypted form. $\mathcal{T}$ is the MFCC transformation, $g$ is a combination of subsequent operations (e.g. a classifier layer) over MFCCs.}
    \label{fig:clientserver}
\end{figure}

One way of performing computations securely is to operate on encrypted data, so that the client data is not at risk during inference, as illustrated in Figure \ref{fig:clientserver}. We assume in this paper that the client is honest \citet{backstrom2023privacy}. This falls into a cryptographic approach known as homomorphic encryption.

A scheme is said to be \textit{fully homomorphic} if it enables the evaluation of any function $f$ over encrypted data. As a result, Fully Homomorphic Encryption (FHE) allows an individual A to compute any function $f$ over some encrypted data $E(x)$ without accessing the \textit{plaintext}, the data $x$, by using the FHE circuit equivalent $\hat{f}$ which operates over encrypted data:
$$
f(x) = E^{-1}\left( \hat{f}\left[ E(x) \right] \right)
$$

In this work, we used the fast fully homomorphic encryption scheme over the torus (TFHE) \citet{TFHE} and its Concrete \citet{Concrete} implementation. We chose TFHE over the more commonly used Cheon-Kim-Kim-Song (CKKS) scheme \citet{ckks} because CKKS can only handle a bounded number of additions and multiplications, therefore it cannot truly handle any function $f$. Moreover, contrary to TFHE, CKKS is approximate by design as it introduces a noise that can corrupt the data \citet{tfhe_vs_ckks}. 


TFHE and CKKS both represent numbers as integers to perform computations on encrypted data. To compile a given function $f$ to an equivalent FHE circuit $\hat{f}$, Concrete computes look-up tables which store all possible intermediate values in computing $\hat{f}$ as integers rather than floating-point values. Concrete also enforces a 16-bit limit over all intermediate values of the circuit in order to keep the FHE duration low. As a result, $f$ must be quantized to $\Tilde{f}$ according to this 16-bit integer limit before being compiled to $\hat{f}$. In the remainder of this work, we refer with a tilde all functions $\Tilde{f}$ or tensors $\Tilde{X}$ that go through quantization, and with a hat all functions $\hat{f}$ that are compiled to be FHE compatible. Note that writing the compiled FHE circuit $\hat{f}$ always entails that its \textit{clear} counterpart $f$ was quantized to $\Tilde{f}$ before compilation.

\subsection{Quantization}\label{section:quantization}


FHE constraints of representing numbers as low-bit integers entails the use of quantization methods which turn high-bit float values into low-bit integers. 
Quantization on $B$ bits associates each element of a continuous or large set (for example, a subset of $\mathbb{R}$) to a smaller, discrete set of size $2^B-1$ (for example, $\llbracket0,  2^{B}-1\rrbracket$). The transformation of a floating value to an integer results in loss of information. However, quantization provides speed gains that are necessary for the compilation of FHE circuits. 

In this work, we used the range-based, affine, uniform quantization \citet{quantization}. Let $[\alpha, \beta]$ be the real range of values represented by floats in the calibration data, and $\llbracket0, 2^B-1\rrbracket$ the quantized range. Each float value $x$ is mapped to a quantized value $\Tilde{x}$ in $\llbracket 0, 2^B-1\rrbracket$ as follows:
\begin{equation}\label{eq:quant}
    \Tilde{x} = \left\lfloor (x - \alpha) \frac{2^B - 1}{\beta - \alpha} \right\rceil,
\end{equation}
where $\lfloor \cdot \rceil$ denotes the rounding to the nearest integer operator. 
Together, $\alpha, \beta$ and $B$ are the quantization parameters which define the quantization operation. The choice of bit width $B$ is a design choice which comes as a trade-off between loss of precision and gains in memory and speed. On the other hand, $\alpha$ and $\beta$ are determined during the \textit{calibration process}, where some calibration data representative of the overall input data distribution is passed to the quantizer to determine quantization parameters. Here, $\alpha$ and $\beta$ are simply taken as the minimum and maximum values in the calibration data.
Thus, the quantization operation defined by $\alpha$ and $\beta$ can be written as $q_{\mathscr{D}, B}$ so that $\Tilde{x} = q_{\mathscr{D}, B}(x)$
where $\mathscr{D}$ is the calibration data. In Equation \eqref{eq:quant}, we used unsigned integers in the quantization range $\llbracket 0, 2^B-1\rrbracket$. In many cases, it can be preferable to use signed integers. In this case, the quantization range becomes $\llbracket -2^{B-1}, 2^{B-1} - 1 \rrbracket$. This is easily performed by substracting $2^{B-1}$ in the calculation of $\Tilde{x}$ in~\eqref{eq:quant}. The operator $q_{\mathscr{D}, B}$ can also be used on data that has already been quantized, for example when a change of bit width is required.

\subsection{Secure audio processing}

To summarize, our goal is to perform secure computation for audio tasks: time-frequency representations from raw audio, audio descriptors and classifiers. We decomposed it as the computation of time-frequency representations denoted by $\mathcal{T}$ and subsequent operations denoted by $g$, which can be the identity function, an audio descriptor or a learned classifier.

Writing as $x$ the raw audio signal, our goal for private processing of audio can be written as:

\begin{equation}\label{def:goal}
    (g \circ \mathcal{T}) (x) \approx E^{-1}(\widehat{g \circ \mathcal{T}}[E(x)]),
\end{equation}
where $\widehat{g \circ \mathcal{T}}$ denotes the FHE circuit equivalent to $g \circ \mathcal{T}$. Note that Equation \eqref{def:goal} is not an equality but an approximation, since the quantization that occurs before compilation leads to a loss of precision in the outputs. 

\section{Approximate quantized signal processing}\label{sec:qasp}

The biggest loss of information to turn a function to its FHE equivalent, in Equation \eqref{def:goal}, lies in the quantization that takes place before compiling \citet{TFHE}. The compilation error can be neglected in comparison to the quantization error. We can consider $E^{-1}(\widehat{g \circ \mathcal{T}}[E(x)]) \approx \widetilde{g \circ \mathcal{T}}(x)$. The challenge is therefore to reduce the quantization error $\Vert \widetilde{g \circ \mathcal{T}}(x) - g \circ \mathcal{T}(x)\Vert$ as much as possible.

This section focuses on the quantization of $\mathcal{T}$. We first set up a general framework for analysis, then introduce approximations of the STFT and explains how they can be used to improve the computation of signal processing with integers with a limited number of bits. An illustration can be found in Figure \ref{fig:STFT}. 

\subsection{Signal processing operations as neural network operations}
\label{sub_section:signal_processing_dl}

The STFT can be seen as a convolution over the signal $x$ with stride $h$, kernels of length $N$, and weights $w(n-mh) \exp\left( \frac{-2j \pi kn}{N}\right)$ (Equation \eqref{def:stft}). We cast the STFT computations as neural network operations \citet{nnaudio} by leveraging convolutional neural networks. Similarly, we wrote Mel filterbanks, mel-frequency cepstrum coefficients (MFCCs) and gammatone filters as neural networks comprised of convolutional layers with fixed kernel weights. This formulation enabled us to use quantization methods with our signal processing formulation which supports customized bit widths for input, output and weights of quantized neural network layers, including convolutional layers. We used Brevitas \citet{brevitas} for the automatic implementation of quantization, transforming $g \circ \mathcal{T}$ into $\widetilde{g \circ \mathcal{T}}$. 


\subsection{Selection of quantization parameters}

\begin{figure*}
\centering
\includegraphics[width=0.95\textwidth]{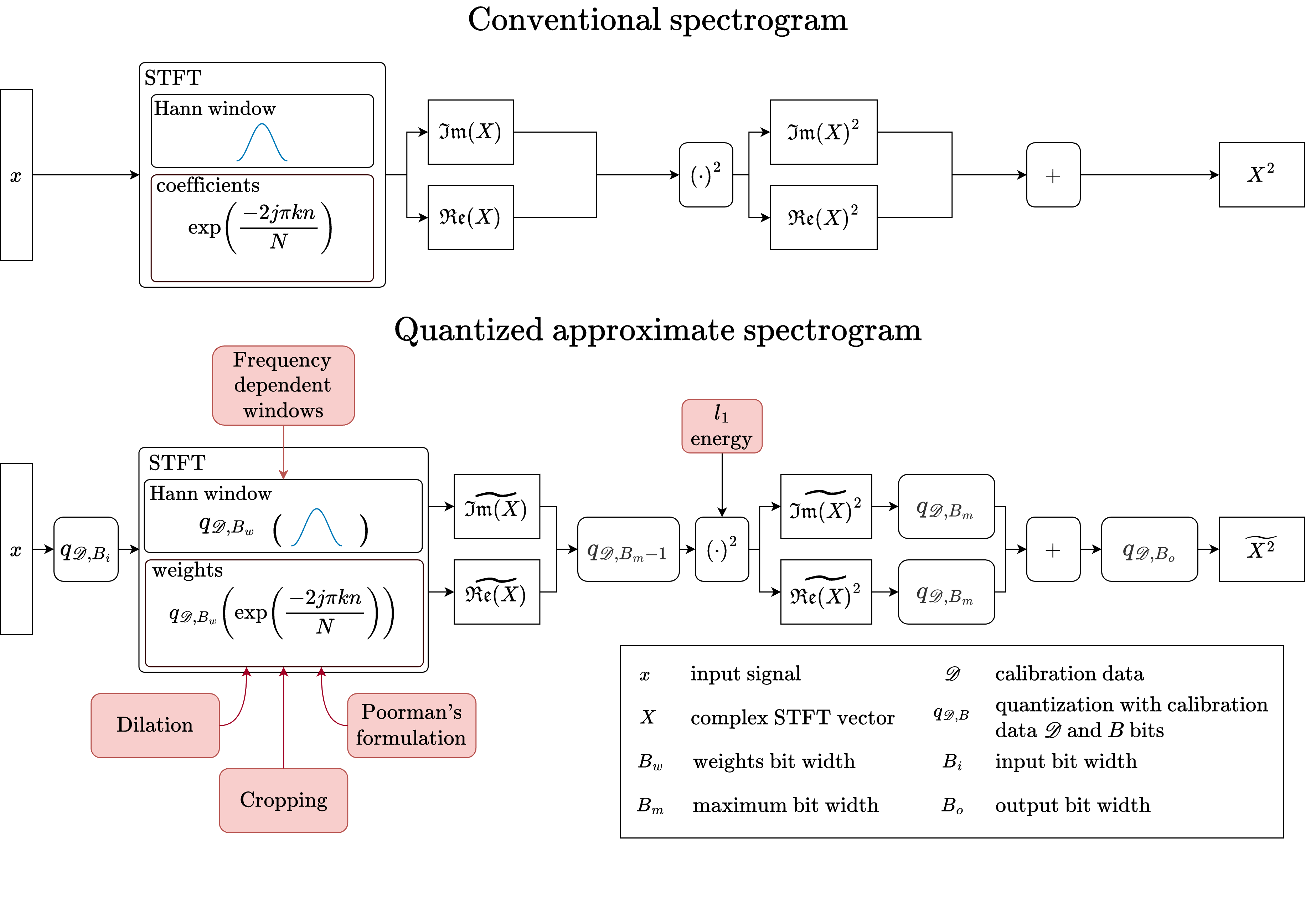}
\caption{Top: conventional computation of the spectrogram, or squared magnitude of the STFT. Bottom: our quantized approximate formulations of the STFT energy.}
\label{fig:STFT}
\end{figure*}

The result of the FHE compilation depends on bit widths parameters controlling the quantization of the model: they are like hyperparameters of the FHE model. For each experiment, we used a grid search to find the set of parameters that leads to the best performances.
Each layer can be controlled with its specific bit width parameters. To simplify the grid search, we used only 4 bit width parameters for the quantization of the time-frequency representations. The repartition of the 4 bit widths parameters in the STFT is shown in Figure \ref{fig:STFT}: parameters $B_i, B_o, B_w, B_m$ respectively control the quantization of the inputs, outputs, convolution weights and intermediary values. We did the same with Mels, MFCCs and gammatone filters intermediate computations. 

\subsection{Quantization of neural networks heuristics}\label{section:framework}

To understand how to reduce the quantization accuracy loss while respecting the 16-bit constraint, it is useful to consider the relationship between: the output bit width of a layer, the bit width of its inputs and weights in the worst case scenario. 

With $m$ and $k$ fixed in Equation \eqref{def:stft}, the STFT computations amount to a scalar product between a convolution kernel and the input signal. 

For a generic kernel $w$ of size $L$ and input vector $x$ of size $L$, we write y the result of the dot product between $w = (w(1), \dots, w(L))$ and $x = (x(1), \dots, x(L))$:

\begin{equation*}
    y = \sum_{l=0}^{L-1} x(l) w(l).
\end{equation*}
Assuming that all $x(l)$ are encoded on $N$ bits and all $w(l)$ are encoded on $M$ bits, also assuming the worst case scenario where every $x(l) = 2^N-1$ and every $w(l) = 2^M-1$, we deduce the number of bits $N'$ needed to encode $y$:

\begin{equation}\label{eq:linear_quantization_constraint}
    N' = \left\lceil \log_2 \left( L (2^{N}-1)(2^M-1) \right) \right\rceil.
\end{equation}
The insight we get from Equation~\eqref{eq:linear_quantization_constraint} is that by reducing $L$ (i.e. introducing zeros in the convolution kernel) or reducing $N$ or $M$ (i.e. quantizing more aggressively the input or the kernel weights), we can decrease the $N'$ value and avoid overflow of the accumulator. Thus introducing zeros in the convolution kernel of the STFT or reducing the number of bits necessary to encode the intermediate values allows us to use more bits in the quantization of either the audio inputs or the rest of the kernel weights. 

Our strategies for strategically inserting zeros in the convolution kernel are inspired by common neural network operations and relate to sparsity structure theory in convolutional neural networks \citet{mao2017exploring}. Two of our approximations, namely \textit{frequency dependent windows} and \textit{dilation}, fall in the category of so-called fine-grained, 0-D sparsity since they affect the kernels at the individual weights level. In contrast, what we introduce as \textit{cropping} refers to more coarse, kernel-level 2-D sparsity.

We introduced approximations $\mathcal{T'}$ of the STFT so that the goal described in Equation \eqref{def:goal} becomes
\begin{equation}\label{def:approxgoal}
    (g \circ \mathcal{T}) (x) \approx E^{-1}(\widehat{g \circ \mathcal{T'}}[E(x)]).
\end{equation}
Using $\mathcal{T'}$ rather than $\mathcal{T}$ could save unnecessary bit computation in the quantization stage and allow an increase in bit widths while still respecting the 16-bit integer constraint. While it would introduce an error between spectrograms $\Vert \mathcal{T}(x)-\mathcal{T'}(x) \Vert > 0$, it could decrease the loss due to quantization so that
\begin{align}
    \Vert \widetilde{g \circ \mathcal{T'}}(x) - g \circ \mathcal{T'}(x) \Vert \leq \Vert \widetilde{g \circ \mathcal{T}}(x) - g \circ \mathcal{T}(x) \Vert.
\end{align}
If this loss decrease proves significant, using the approximate audio transformation $\mathcal{T'}$ could lead to FHE computations that are more accurate to their clear counterparts. However, it is computationally expensive to compute this loss theoretically for any circuits as it is partially data-dependent. Therefore it is hard to know which approximations can lead to better performance. 

Optimizing bit usage in the STFT computation can increase the accuracy for the other time-frequency operators (Mel filterbanks, MFCCs, gammatone filters) because they are all based on the STFT. 
The remainder of this section describes the approximations of the STFT that we use.

\begin{figure*}
\centering
\begin{subfigure}{0.24\linewidth}
    \includegraphics[width=\linewidth]{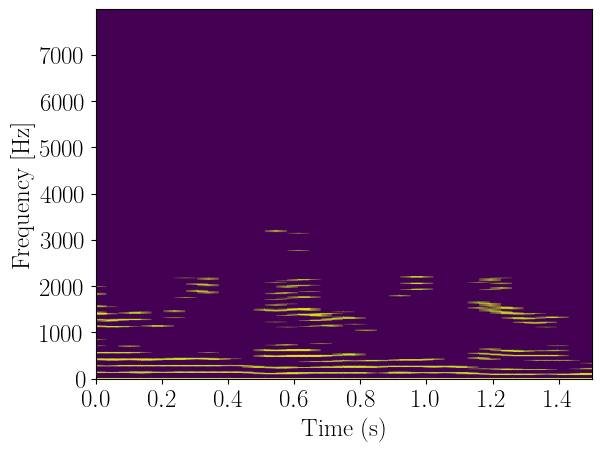}
    \caption{STFT with FHE}
\end{subfigure}
\begin{subfigure}{0.24\linewidth}
    \includegraphics[width=\linewidth]{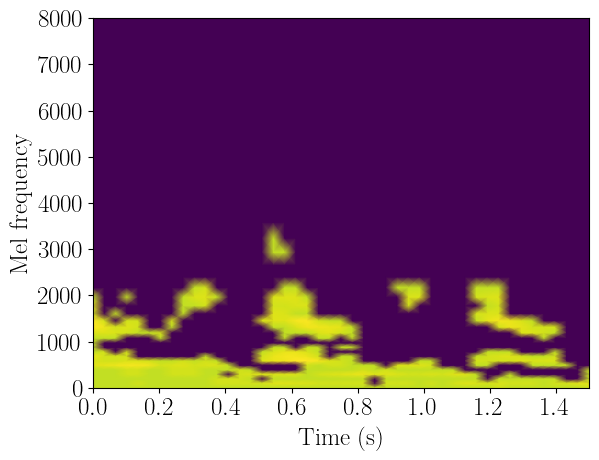}
    \caption{Mel with FHE}
\end{subfigure}
\begin{subfigure}{0.23\linewidth}
    \includegraphics[width=\linewidth]{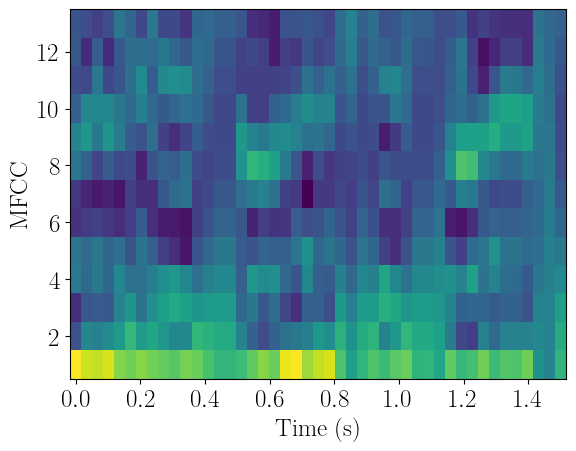}
    \caption{MFCC with FHE}
\end{subfigure}
\begin{subfigure}{0.24\linewidth}
    \includegraphics[width=\linewidth]{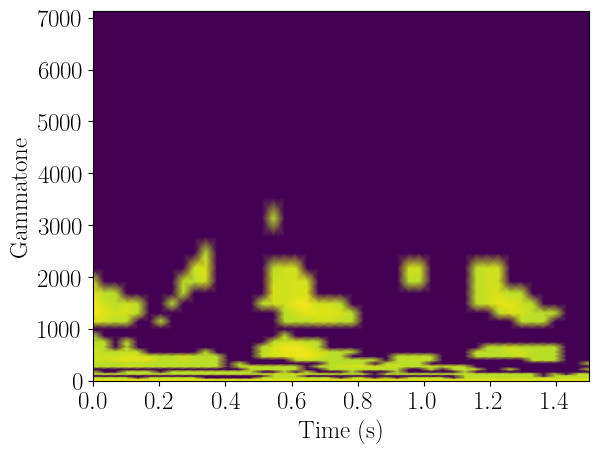}
    \caption{Gammatone with FHE}
\end{subfigure}
\begin{subfigure}{0.24\linewidth}
    \includegraphics[width=\linewidth]{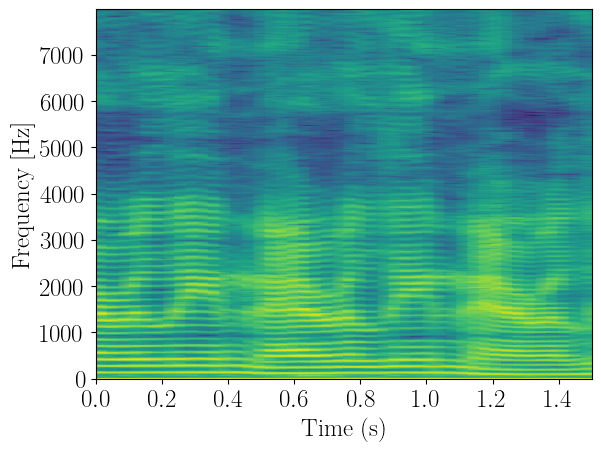}
    \caption{Clear STFT}
\end{subfigure}
\begin{subfigure}{0.24\linewidth}
    \includegraphics[width=\linewidth]{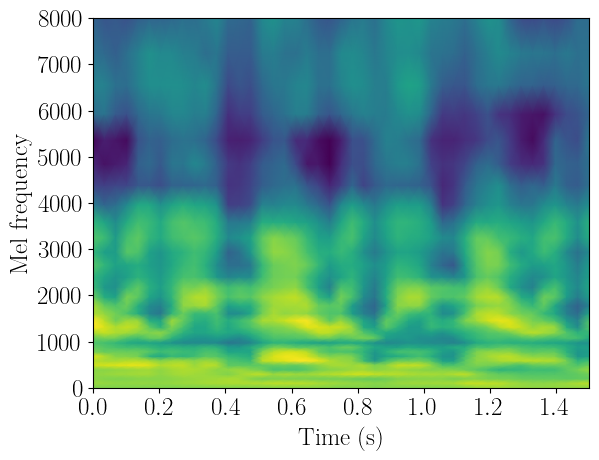}
    \caption{Clear Mel}
\end{subfigure}
\begin{subfigure}{0.23\linewidth}
    \includegraphics[width=\linewidth]{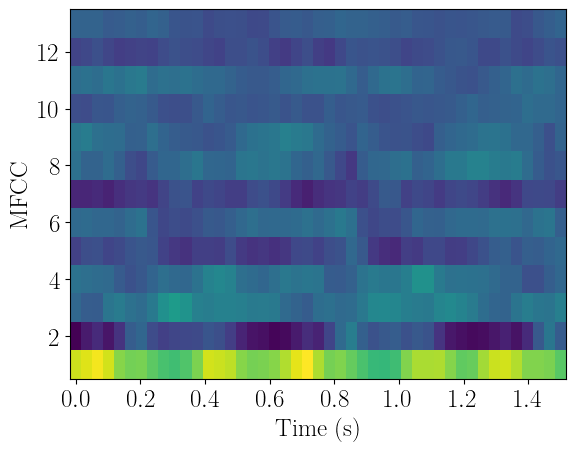}
    \caption{Clear MFCC}
\end{subfigure}
\begin{subfigure}{0.24\linewidth}
    \includegraphics[width=\linewidth]{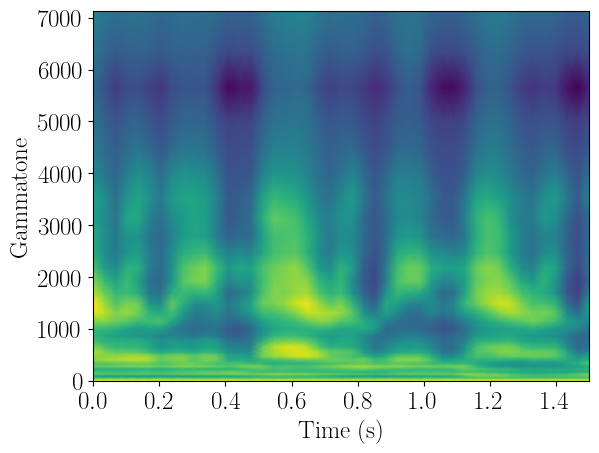}
    \caption{Clear Gammatone}
\end{subfigure}
        
\caption{Comparison of log-scale spectrograms with FHE-friendly transformation corresponding to lowest Euclidean distance, of a given audio from the Vocalset dataset. }
\label{fig:SpectrogramVocalset}
\end{figure*}
\subsection{Convolution dilation}\label{sub_section:dilation}

The first approximation is based on the premise that downsampling can lead to a sparse STFT convolution kernel without compromising the structural integrity of the signal.
This common approximation uses dilated convolutions \citet{dilatedconv} and applies them to the STFT formulation as a convolution operator. It consists in skipping samples of the signal at a chosen rate in the computation of the STFT. With a dilation of factor $d$, one sample is used every $d$ samples. We use the Shannon-Nyquist sampling theorem, which states that the sample rate of a signal must be at least twice the bandwidth of the signal to avoid aliasing, as a heuristic to derive the following maximum dilation rate in each frequency bin:

\begin{align*}
    d_k &= \left\lfloor\frac{f_s}{2f_k}  \right\rfloor = \left\lfloor \frac{N}{2(k+1)f_s} \cdot f_s  \right\rfloor= \left\lfloor \frac{N}{2(k+1)}  \right\rfloor,
\end{align*}
with $f_s$ the sample rate and $f_k$ the upper frequency of the $k$-th frequency bin.

The dilated STFT approximation is defined as:
\begin{equation}\label{approx:dilation}
    X^{(d)}(m, k) := \sum_{n=0}^{M-1}x(n)w(n-mh) e^{{- \frac {2j\pi kn}{N}}} 1_{n \equiv 0[\min(d, d_k)]}.
\end{equation}
where
\begin{align*}
1_{n \equiv 0[\min(d, d_k)]} &= \begin{cases} 1 & \mbox{if $n$ is a multiple of $\min(d, d_k)$ }\\
0 & \mbox{otherwise.}\end{cases}
    \end{align*}
From Equation \eqref{approx:dilation}, the error formulations reads as:
\begin{equation}
    \label{dilation_error1}
    \left|X(m, k) - X^{(d)}(m, k)\right| = \left|\sum_{n=0}^{M-1} x(n)w(n-mh)
    e^{-\frac{2ik\pi n}{N}}\left( 1 - 1_{n \equiv 0[\min(d, d_k)]}\right)\right|.
\end{equation}

From Equation \eqref{dilation_error1} it is clear that when $d$ increases, more noise from other frequency bins is injected. This is especially the case for lower frequencies as they correspond to a higher maximum dilation rate $d_k$.

In particular, the STFT approximation with maximal frequency-dependent dilation can be written as  $X^{dil} := X^{(d_k)}$.

\subsection{Frequency-dependent windows}

In the conventional STFT, a uniform window is applied across all frequency bins, which makes the window width completely frequency independent. The resolution in time and frequency strongly relates to the window width through the well-known time-frequency tradeoff. Intuitively, a signal has to be observed for a certain amount of time (the window has to be a certain width) to witness at least one oscillation of the signal and get an idea of its frequency. As a result, higher time resolution leads to lower frequency resolution, and vice-versa.

At higher frequencies, less time is needed to observe signal oscillations. Variable window transforms such as the Constant-Q Transform (CQT) \citet{cqt} make use of this fact and introduce narrower windows for higher frequencies in order to optimize the time-frequency tradeoff at each frequency bin. We took inspiration from the CQT to define frequency-dependent window widths, with padding applied as necessary to maintain frame length (i.e. kernel size) consistency. 

In this section, we define the frequency-dependent variable window as:\\ $N^{fd}(k) = \min\left(N, N_{min}\frac{f_{\max}}{f_k}\right)$, with $f_{\max}$ the maximum frequency of the spectrogram, $N$ the highest window width we choose to allow and $N_{\min}$ the smallest window width. $N$ and $N_{\min}$ are parameters to choose. From this formula, it is clear that at the lowest frequencies, the window width is equal to $N$, and that at the highest frequency, it is equal to $N_{\min}$. Therefore, the corresponding approximate STFT formulation can be written as:
\begin{align*}
X^{fd}(m, k) &:= \sum_{n=0}^{M-1}x(n)w_k(n-mh) e^{- \frac {2j\pi kn}{N}},\\
\text{with }w_k(n) &= \frac{1}{2}\left(1 - \cos\frac{2\pi n}{N^{fd}(k) - 1}\right) 1_{\frac{N - N^{fd}(k)}{2} \leq n \leq \frac{N + N^{fd}(k)}{2}}.
\end{align*}
The STFT error formulation is:
\begin{align*}
    |X(m, k)& - X^{fd}(m, k)| = \left|\sum_{n=0}^{M-1} x(n)e^{-\frac{2ik\pi n}{N}}\left(w- w_k\right) (n-mh)\right|,
\end{align*} 
where
\begin{align*}
    (w - w_k)(n-mh) &= \begin{cases} \frac{1}{2}\left(\cos \frac{2\pi n}{N - 1}- \cos \frac{2\pi n}{N^{fd}(k) - 1}\right) & \mbox{if } \frac{N - N^{fd}(k)}{2} \leq n \leq \frac{N + N^{fd}(k)}{2} \\
    -w(n-mh) & \mbox{otherwise.}\end{cases}
\end{align*} 
This error formulation distinguishes two types of approximation errors: the padding error $-w(n-mh)$ due to the added zeros in order to keep a consistent frame length, and the narrowing error $\frac{1}{2}\left(\cos \frac{2\pi n}{N - 1}- \cos \frac{2\pi n}{N(k) - 1}\right)$ due to the narrower window functions, which increase resolution in time but decrease resolution in frequency due to the time-frequency tradeoff. Our intuition is that the windows only get narrow at high frequencies, and the loss of resolution is not too drastic. 

\subsection{Poorman's formulation}

We propose to use the poorman's DFT \citet{Poorman} formulation of the DFT, a reduction of the frequency granularity. The idea is to project each complex integral $e^{-2j \pi n \frac{k}{N}}$ on the set $\left\{ 1, -1, j, -j\right\}$. With that approximation, the only operations that act on the input signal are changes in sign and/or conjugation. When applying this to our STFT computation, it means that more bits can be used to encode the window function, the input or the next layers. This approximation consists not in the insertion of zeros but rather in the targeted quantization of part of the STFT computation. 

To generalize the poorman's process, we proposed to project each complex integral on a bigger set $\left\{ e^{-2j \pi \frac{l}{L}}, l < L \right\}$ where $L$ is an integer. The original formulation corresponds to $L = 4$. We denote by $p_L: \mathbb{U} \rightarrow \mathbb{U}_L$ the projection from the unit circle to $\mathbb{U}_L$ and define the poorman's STFT as
\begin{equation}
    X^L(m, k) := \sum_{n = 0}^{M-1} x(n)w(n-mh) p_L\left(e^{{- \frac {2j\pi kn}{N}}}\right).
\end{equation}
We derived the following upper bound on the STFT error in appendix \ref{demo:poorman}:
\begin{equation*}
    \left|X(m, k) - X^{L}(m, k)\right| \leq 2\left|\sin\frac{\pi}{2L}\right| \sqrt{\sum_{i=0}^{M-1}(x(i)w(i-mh))^2}.
\end{equation*}
We obtained a $k$-independent upper bound for the error, that vanishes to zero when $L$ goes to infinity.

\subsection{$l_1$ energy}

The computation of the module of the $X$ vector implies the squaring of its real and imaginary parts. This operation is costly: in the worst case scenario, the number of bits needed to represent the output is twice the number of bits used for the input. A workaround is to approximate the spectral density by replacing $|X|^2$ with $\|X\|_{l_1} :=|\Re(X)| + |\Im(X)|$, which only leads to one additional bit to represent the output in the worst case scenario. In this case the error is simply
$$
\left||X|^2- \|X\|_{l_1}\right| = \left|\Re(X)^2 + \Im(X)^2 - |\Re(X)| - |\Im(X)|\right|
$$


\subsection{Cropping}\label{sub_section:cropping}




Another way to insert zeros in the STFT kernel is to set all STFT coefficients to 0 in a certain range of frequencies, i.e. to compute the STFT in a narrower range of frequencies. 

We defined the cropping frequencies approximation, which consists in setting all STFT coefficients $\exp\left({- \frac {2j\pi kn}{N}}\right)$ to 0 where $f_k \notin [f_{min}, f_{max}]$, with $f_{min}$ and $f_{max}$ hyperparameters to choose. Thus the STFT error at each time $m$ and frequency $k$ is simply
$$
\left|X(m, k) - X^{crop}(m, k)\right| =\left|X(m, k)\right| (1_{f_k  > f_{max}} + 1_{f_k < f_{min}})
$$
Our heuristic is that in voice, most of the important information contained in the signal comes from the lower frequencies, so we chose as parameters $f_{min} = 0$ and $f_{max} = 1000$ Hz. 

\section{Experimental setup}\label{section:experiments}

Our goal is to compute quantized, FHE-compatible, approximate time-frequency representations securely with $E^{-1}(\widehat{g \circ \mathcal{T'}}[E(x)])$ (Equation \eqref{def:approxgoal}). In Section~\ref{sec:qasp} we focused on the quantization of $\mathcal{T}$. In this section, we present our experiments with different $g$ functions. In our experiments, we evaluated the fidelity of the secure computation $E^{-1}(\widehat{g \circ \mathcal{T'}}[E(x)])$ to its clear counterpart $(g \circ \mathcal{T})(x)$, where $g$ is the identity function, an audio descriptor computing statistics over the spectrograms, or a learned classifier. Additionally, we compared the different approximations of the STFT and determined whether they provide an increase in accuracy compared to the baseline.

To assess the quality of the secure computation to its clear counterpart, we considered both intrinsic (based on raw spectrograms) and extrinsic (based on downstream tasks such as audio descriptor computation or classification accuracy) performance. We considered the four audio transformations mentioned in Section \ref{section:context}, all based on the standard STFT or its approximate counterpart, and multiple audio descriptors. 

\subsection{Metrics}

\subsubsection{Intrinsic metric}

We used an intrinsic metric to measure the distance between FHE and clear spectrograms ($g=Id$). Specifically, we used the 2D Euclidean distance between spectrograms normalized by their L2 norm, to emphasize the repartition of the energy throughout the spectrogram rather than the raw energy values.

\subsubsection{Audio descriptors}

We computed several audio descriptors in FHE over secure spectrograms:
\begin{itemize}
    \item the average of the standard deviation over time of the energy for each frequency bin, for Mel or Gammatone features,
    \item the mean over time of the root-mean-square (RMS) values, computed over the STFT spectrogram
    \item the standard deviation over time of the RMS values, computed over the STFT spectrogram.
\end{itemize}

Computing audio descriptors over spectrograms indicates how much of the spectrogram information necessary to compute the audio descriptor was preserved by FHE. Additionally, it provides a harsher quantization challenge than the previous task, since more operations are performed, on top of time-frequency representations. We computed the mean of standard deviations for Mel filterbanks and Gammatones, quantifying potential source disturbances in vowels \citet{quatieri2012vocal, riad2020vocal}; and we computed RMS statistics as commonly used in various paralinguistic tasks \citet{eyben2010opensmile}.
The compiled model outputs all 4 audio descriptors in one tensor, so they are required to be quantized with the same parameters. Since the scales of values of the audio descriptors can be different, we normalized before concatenation and the common quantization of the descriptors to avoid unnecessary loss of precision. We computed constants during the calibration process and used them at inference time to simulate regular normalization.

For a given $\mathcal{T}'$ (conventional or approximation of the STFT), we selected best bit width parameters $B_i, B_o, B_w$ and $B_m$ with the Pearson correlation coefficient between the outputs of the FHE and clear descriptors. To compare approximations against each other, we performed an in-depth evaluation of false discovery and missing discovery rates. We compared the results of statistical tests performed using clear and FHE outputs. Specifically, we computed the audio descriptors for all audio classes in both datasets using both FHE and clear computation. Then, we performed a Mann-Whitney U test on the audio descriptors from each pair of audio classes in both FHE and clear computation. A high-fidelity FHE audio descriptor is expected to yield p-values for these statistical tests that are similar to the p-values obtained with the audio descriptor in clear. We defined a true positive as a pair of audio classes where the FHE computation maintains the statistical test separation ($p < 0.05$) observed in the clear computation. Conversely, a false negative occurs when the FHE computation fails to preserve this separation. True negatives and false positives were defined similarly. By calculating the error rate (sum of false discovery and missed discovery) introduced by FHE, we gauged the accuracy of the FHE audio descriptors. We illustrated this evaluation for a given marker in Figure \ref{fig:pvals_scatter} left, and a specific comparison of Clear and FHE statistical test on Figure \ref{fig:pvals_scatter} right. 

\subsubsection{Classification}

Finally, we performed gender and vocal exercise classification. We trained a convolutional neural network (CNN) to classify audios either in the clear or in FHE, and compared the obtained accuracies. We used a CNN architecture containing two convolutional blocks with respectively 8 and 16 filters, ReLU activations and batch normalizations, a max-pooling layer and two fully connected layers separated by ReLU activations. We used a batch size of $32$ for gender classification and $16$ for vocal exercise classification, and an Adam optimizer with a learning rate of $10^{-3}$ for $10$ to $20$ epochs. We performed quantization-aware training, meaning we trained the quantized network directly instead of performing post-training quantization. 
\subsection{Datasets}

We used two real-world audio datasets for the experiments: VocalSet  \citet{Vocalset} and OxVoc Sounds databases \citet{oxvoc}. 

The VocalSet database consists of 10 hours of 11 male and 9 female professional singers performing 4 vocal exercises each with different variations. We considered gender classes and vocal exercise classes (arpeggios, scales, long tones and song excerpts) for the classification task. For replication of statistical tests with audio descriptors, we considered more fine-grained classes: we performed statistical tests over all pairs of variations inside each vocal exercise, which amounts to 144 pairs of audio classes (See Annex for full list of variations for all 4 vocal exercises). 

The OxVoc Sounds database consists of 173 spontaneous non-verbal vocal sounds from infants, adults and domestic animals. Human sounds comprise natural expressions of happy, neutral and sad emotions. For adults, sounds are further subdivided by the gender of the speaker. We did not consider this dataset for the CNN classification task, as its size is not sufficient. For the task of replicating statistical tests with audio descriptors, we considered pairs of gender categories in each adult emotion class, pairs of overall speaker categories (infant, adult or animal), and speaker subcategories (happy, sad and neutral for humans, cats and dogs for animals), which amount to 27 pairs of audio classes.

For computing FHE spectrograms and audio descriptors, we split each dataset into calibration and evaluation sets by selecting 10\% of the dataset for calibration and the remaining 90\% for evaluation. We used stratified sampling to make sure the class proportions remained consistent with those of the entire dataset. For classification, we also used stratified sampling to split the VocalSet dataset into train and test sets, using 80\% of the data to train and 20\% to test.

\section{Results and discussions}\label{sec:results}

Our results are presented in increasing order of interest towards recent audio processing tasks. First, we examined intrinsic metrics over FHE spectrograms. Then, we examined the statistical properties of FHE audio descriptors, meaning that the compiled FHE model outputs not only one audio descriptor but a vector composed of all audio descriptors. This necessitated a deeper FHE circuit and put harsher constraints on the quantization of the model. Finally, we built and evaluated classifiers to output classes directly from private audio.
\begin{figure*}[h!]
    \centering
    \includegraphics[width=\linewidth]{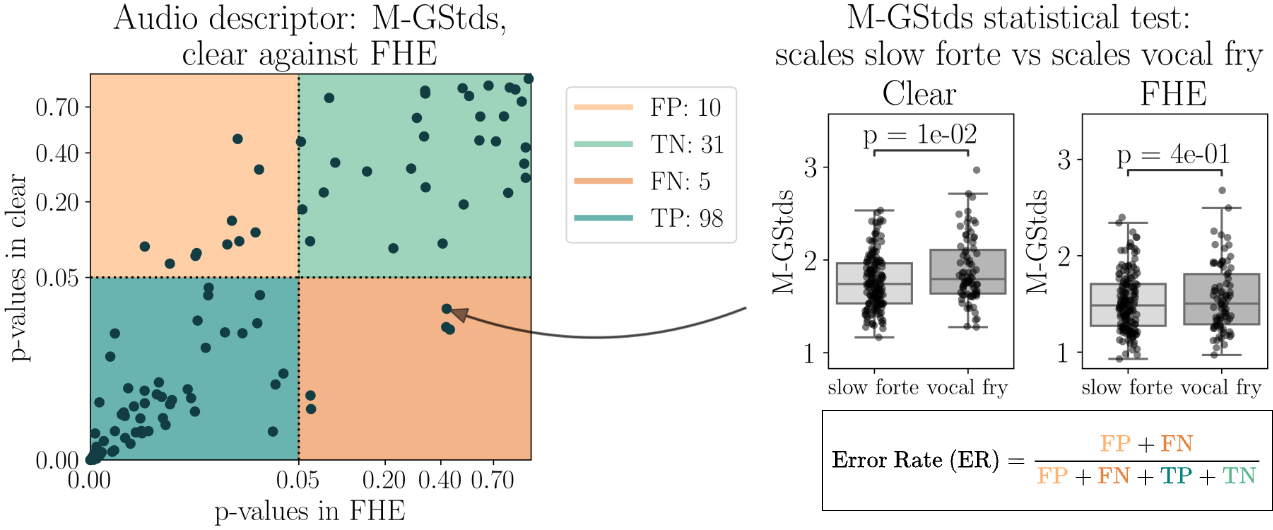}
    \caption{Example of a scatter plot of p-values computed in clear or in FHE with the conventional STFT formulation. Only the mean of gammatone standard deviations (M-GStds) audio descriptor is plotted. The data is from VocalSet. The arrow points to statistical tests in clear and FHE for the pair of audio classes ("scales slow forte", "scales vocal fry"). }
    \label{fig:pvals_scatter}
\end{figure*}

\subsection{Intrinsic metric results}\label{section:intrinsic}

We displayed in Figure \ref{fig:SpectrogramVocalset} all 4 time-frequency representations for audio extracted from the Vocalset dataset, both in clear and with FHE using the conventional STFT. Qualitatively, although the harmonics are lost above 4000 Hz, some of the patterns in lower frequencies are preserved in the spectrograms. We also showed in Table \ref{table:euclidean} the means, standard deviations and ranges of the average 2D Euclidean distance between the normalized clear and FHE spectrograms for each dataset. 

\begin{table}[!ht]
\centering
\begin{tabularx}{\linewidth}{@{}l *{4}{>{\centering\arraybackslash}X}@{}}
\toprule
\multirow{2}{*}{\begin{tabular}{@{}l@{}}Audio \\features\end{tabular}} 
& \multicolumn{2}{c}{VocalSet} 
& \multicolumn{2}{c}{OxVoc} \\
\cmidrule(lr){2-3}\cmidrule(lr){4-5}
& Mean \(\pm\) Std & Range & Mean \(\pm\) Std & Range \\
\midrule
STFT & $0.13 \pm 0.14$ & $[0.00, 1.38]$ & $0.13 \pm 0.17$ & $[0.02, 1.37]$ \\
Mel & $0.18 \pm 0.19$ & $[0.00, 1.38]$ & $0.13 \pm 0.16$ & $[0.02, 1.21]$ \\
MFCC & $0.78 \pm 0.16$ & $[0.31, 1.45]$ & $0.85 \pm 0.17$ & $[0.41, 1.53]$ \\
GammaT & $0.18 \pm 0.19$ & $[0.01, 1.41]$ & $0.15 \pm 0.19$ & $[0.03, 1.21]$ \\
\bottomrule
\end{tabularx}
\caption{2D Euclidean distances including mean \(\pm\) standard deviation and range for either VocalSet or OxVoc Sounds datasets, between normalized clear and FHE spectrograms. GammaT stands for Gammatones.}
\label{table:euclidean}
\end{table}

As shown qualitatively in Figure \ref{fig:SpectrogramVocalset}, the MFCC time-frequency representation suffered the most from the quantization necessary to compile the model, as it showed the highest 2D Euclidean mean distance on both datasets. On the other hand, the three other audio transformations showed relatively low distances consistent across both datasets. Yet, for STFT, Mel filterbanks and Gammatones, there were outliers (maximum values above 1) in both datasets even though the mean and standard deviation remain low under 20\%. 

\subsection{FHE audio descriptors}
\begin{table*}[ht]
\centering
\begin{tabularx}{\textwidth}{@{\extracolsep{\fill}} c *{4}{>{\centering\arraybackslash}X}@{}}
\toprule
\multirow{2}{*}{\begin{tabular}{@{}c@{}}STFT\\formulation\end{tabular}}  
& \multicolumn{2}{c}{Mean Gammatone stds}  
& \multicolumn{2}{c}{Mean Mel stds}  \\
\cmidrule(lr){2-3}\cmidrule(lr){4-5}
 & \begin{tabular}{@{}c@{}}VocalSet \\ ($n = 144$)\end{tabular}  
 & \begin{tabular}{@{}c@{}}OxVoc \\ ($n = 27$)\end{tabular}  
 & \begin{tabular}{@{}c@{}}VocalSet \\ ($n = 144$)\end{tabular}  
 & \begin{tabular}{@{}c@{}}OxVoc \\ ($n = 27$)\end{tabular}  \\
\midrule
Conventional STFT & 10.4 (15) & 7.4 (2) & 9.0 (13) & 7.4 (2) \\
\midrule
Poorman $L = 4$ & 9.0 (13) & 3.7 (1) & 8.3 (12) & 3.7 (1) \\
Poorman $L = 6$ & 11.1 (16) & 3.7 (1) & 9.0 (13) & \textbf{0.0 (0)} \\
Poorman $L = 8$ & 12.5 (18) & \textbf{0.0 (0)} & 8.3 (12) & \textbf{0.0 (0)} \\
\midrule
Dilation $d = 2$ & 13.9 (20) & 3.7 (1) & 15.3 (22) & \textbf{0.0 (0)} \\
Dilation $d = 3$ & \textbf{6.9 (10)} & 11.1 (3) & 10.4 (15) & 14.8 (4) \\
Dilation $d = 4$ & 8.3 (12) & \textbf{0.0 (0)} & 8.3 (12) & 3.7 (1) \\
Dilation $d = 5$ & 7.6 (11) & 7.4 (2) & \textbf{7.6 (11)} & 7.4 (2) \\
Dilation $d=d_k$ & 13.9 (20) & 29.6 (8) & 16.7 (24) & 33.3 (9) \\
\midrule
FreqAW & 11.8 (17) & 25.9 (7) & 10.4 (15) & 22.2 (6) \\
\midrule
L1 & 29.2 (42) & 37.0 (10) & 24.3 (35) & 44.4 (12) \\
\midrule
Cropping & 10.4 (15) & 3.7 (1) & 9.0 (13) & 37.0 (10) \\
\bottomrule
\end{tabularx}
\caption{ Error rates for statistical tests with FHE for Mean Gammatone and Mean Mel standard deviations on VocalSet and OxVoc Sounds datasets, over different approximations. $n$ is the total number of statistical tests for pairs of audio classes. The absolute sum of false positives and false negatives is in parentheses. The best approach for each marker and dataset is \textbf{bolded}. Cropping was performed with $1$kHz. $d = d_k$ stands for choosing the maximum dilation rate $d_k$ at each frequency bin $k$. FreqAW stands for frequency-adapted windows, performed with $N_{min} = 80$. }
\label{table:mean_gammatone_mel}
\end{table*}

\begin{table*}[!ht]
\centering
\begin{tabularx}{\textwidth}{@{\extracolsep{\fill}} c *{4}{>{\centering\arraybackslash}X}@{}}
\toprule
\multirow{2}{*}{\begin{tabular}{@{}c@{}}STFT\\formulation\end{tabular}}  
& \multicolumn{2}{c}{Std RMS}  
& \multicolumn{2}{c}{Mean RMS}  \\
\cmidrule(lr){2-3}\cmidrule(lr){4-5}
 & \begin{tabular}{@{}c@{}}VocalSet \\ ($n = 144$)\end{tabular}  
 & \begin{tabular}{@{}c@{}}OxVoc \\ ($n = 27$)\end{tabular}  
 & \begin{tabular}{@{}c@{}}VocalSet \\ ($n = 144$)\end{tabular}  
 & \begin{tabular}{@{}c@{}}OxVoc \\ ($n = 27$)\end{tabular}  \\
\midrule
Conventional STFT & 2.8 (4) & 11.1 (3) & 1.4 (2) & \textbf{0.0 (0)} \\
\midrule
Poorman $L = 4$ & \textbf{1.4 (2)} & 7.4 (2) & \textbf{0.7 (1)} & \textbf{0.0 (0)} \\
Poorman $L = 6$ & 2.1 (3) & 18.5 (5) & 1.4 (2) & 3.7 (1) \\
Poorman $L = 8$ & \textbf{1.4 (2)} & 11.1 (3) & \textbf{0.7 (1)} & \textbf{0.0 (0)} \\
\midrule
Dilation $d = 2$ & \textbf{1.4 (2)} & 7.4 (2) & \textbf{0.7 (1)} & \textbf{0.0 (0)} \\
Dilation $d = 3$ & 9.7 (14) & 7.4 (2) & 6.3 (9) & \textbf{0.0 (0)} \\
Dilation $d = 4$ & 2.1 (3) & \textbf{3.7 (1)} & \textbf{0.7 (1)} & \textbf{0.0 (0)} \\
Dilation $d = 5$ & 3.5 (5) & 7.4 (2) & 4.9 (7) & 3.7 (1) \\
Dilation $d=d_k$ & 16.7 (24) & 37.0 (10) & 26.4 (38) & 18.5 (5) \\
\midrule
FreqAW & 6.9 (10) & 29.6 (8) & 7.6 (11) & 3.7 (1) \\
\midrule
L1 & 31.3 (45) & 25.9 (7) & 22.9 (33) & \textbf{0.0 (0)} \\
\midrule
Cropping & 2.8 (4) & 7.4 (2) & 1.4 (2) & \textbf{0.0 (0)} \\
\bottomrule
\end{tabularx}
\caption{ Error rates for statistical tests with FHE for Std RMS and Mean RMS on VocalSet and OxVoc Sounds datasets, over different approximations. $n$ is the total number of statistical tests for pairs of audio classes. The absolute sum of false positives and false negatives is in parentheses. The best approach for each marker and dataset is \textbf{bolded}. Cropping was performed with $1$kHz. $d = d_k$ stands for choosing the maximum dilation rate $d_k$ at each frequency bin $k$. FreqAW stands for frequency-adapted windows, performed with $N_{min} = 80$. .}
\label{table:std_mean_rms}
\end{table*}

Figure \ref{fig:pvals_scatter} illustrates our results for the mean of gammatone standard deviations (M-GStds) audio descriptor with the conventional STFT approach, and shows the statistical test results for a false negative example. We showed in Table \ref{table:mean_gammatone_mel} and in Table \ref{table:std_mean_rms} the full examination of error rates of the conventional STFT with FHE, in comparison to our introduced approximation. We displayed the total number and error rates (sum of false positives and false negatives) for the VocalSet Singing dataset and the OxVoc Sounds dataset. 

We found out that (Table \ref{table:mean_gammatone_mel}, Table \ref{table:std_mean_rms}) that the conventional STFT approach is making some errors but remain low, as it only lead to mean error rates of $5.9\%$ and $6.5\%$ for VocalSet and OxVoc respectively. Secondly, the poorman's transform approach improved upon the baseline in both datasets in its original formulation ($L=4$ quantized angles) by achieving mean error rates of $4.9\%$ and $3.7\%$ for VocalSet and OxVoc respectively. The dilation approach also outperformed the baseline conventional STFT, especially with dilation factor $4$ which turned out to be the best approach across both datasets (mean error rates $4.7\%$ and $1.9\%$ for VocalSet and OxVoc respectively). In contrast, other approaches such as cropping, frequency-adapted windows or L1 energy lead to worse results.

\subsection{Classification in FHE}\label{classification}

Finally, our spectrogram-based classification results are presented in Table \ref{table:gender_classif} and in Table \ref{table:vocal_classif}.
\begin{table}[ht]
\centering
\begin{tabularx}{\linewidth}{@{}l *{4}{>{\centering\arraybackslash}X}@{}}
\toprule
Features & Clear & Conventional & Poorman & Dilation \\
\midrule
STFT  & $0.93 \pm 0.01$ & \boldsymbol{$0.89 \pm 0.01$} & $0.82 \pm 0.02$ & $0.88 \pm 0.01$ \\
Mel   & $0.93 \pm 0.01$ & \boldsymbol{$0.86 \pm 0.02$} & $0.84 \pm 0.01$ & $0.82 \pm 0.01$ \\
MFCC  & $0.90 \pm 0.02$ & \boldsymbol{$0.86 \pm 0.04$} & $0.83 \pm 0.01$ & $0.79 \pm 0.01$ \\
GammaT & $0.94 \pm 0.01$ & \boldsymbol{$0.89 \pm 0.02$} & $0.83 \pm 0.02$ & $0.82 \pm 0.01$ \\
\midrule
Chance & \multicolumn{4}{c}{$0.52$} \\
\bottomrule
\end{tabularx}
\caption{Mean and standard deviation of the test set accuracies for binary gender classification obtained with Clear, FHE, Poorman, or Dilation classifiers, across 5 runs with different seeds. Experiments are done on VocalSet. GammaT stands for Gammatones. The best approach for each audio representation is \textbf{bolded}. The "Chance" classifier is a dummy classifier that always predicts the majority class.}
\label{table:gender_classif}
\end{table}
\begin{table}[ht]
\centering
\begin{tabularx}{\linewidth}{@{}l *{4}{>{\centering\arraybackslash}X}@{}}
\toprule
Features & Clear & Conventional & Poorman & Dilation \\
\midrule
STFT  & $0.62 \pm 0.02$ & $0.54 \pm 0.03$ & $0.50 \pm 0.04$ & \boldsymbol{$0.59 \pm 0.01$} \\
Mel   & $0.63 \pm 0.01$ & $0.41 \pm 0.01$ & $0.42 \pm 0.01$ & \boldsymbol{$0.50 \pm 0.02$} \\
MFCC  & $0.56 \pm 0.01$ & $0.53 \pm 0.00$ & $0.50 \pm 0.01$ & \boldsymbol{$0.54 \pm 0.01$} \\
GammaT & $0.68 \pm 0.01$ & \boldsymbol{$0.63 \pm 0.01$} & \boldsymbol{$0.63 \pm 0.01$} & $0.62 \pm 0.02$ \\
\midrule
Chance & \multicolumn{4}{c}{$0.39$} \\
\bottomrule
\end{tabularx}
\caption{Mean and standard deviation of the test set accuracies for 4-class vocal exercise classification obtained with Clear, FHE, Poorman, or Dilation classifiers, across 5 runs with different seeds. Experiments are done on VocalSet. GammaT stands for Gammatones. The best approach for each audio representation is \textbf{bolded}. The "Chance" classifier is a dummy classifier that always predicts the majority class.}
\label{table:vocal_classif}
\end{table}

Binary gender classification accuracies in FHE stayed close to the accuracies in clear audio, ranging from $0.86$ to $0.89$ in FHE compared to $0.90$ to $0.94$ in clear audio. The performance was less sensitive to audio representation. On the other hand, for the more challenging 4-class vocal exercise classification, the audio representation choice mattered more and our approximations helped to get closer to the clear computation. Dilation for the vocal exercise classification was the best for STFT, mel filterbanks and MFCCs. Overall though, these results still suggest that FHE time-frequency representations effectively preserve some, if not much of the information needed for both speaker gender and vocal exercise classification. Approximations could be useful for some computational tasks.

\subsection{Post-hoc analysis}

\begin{figure}[!ht]
    \centering
    \includegraphics[width=\linewidth]{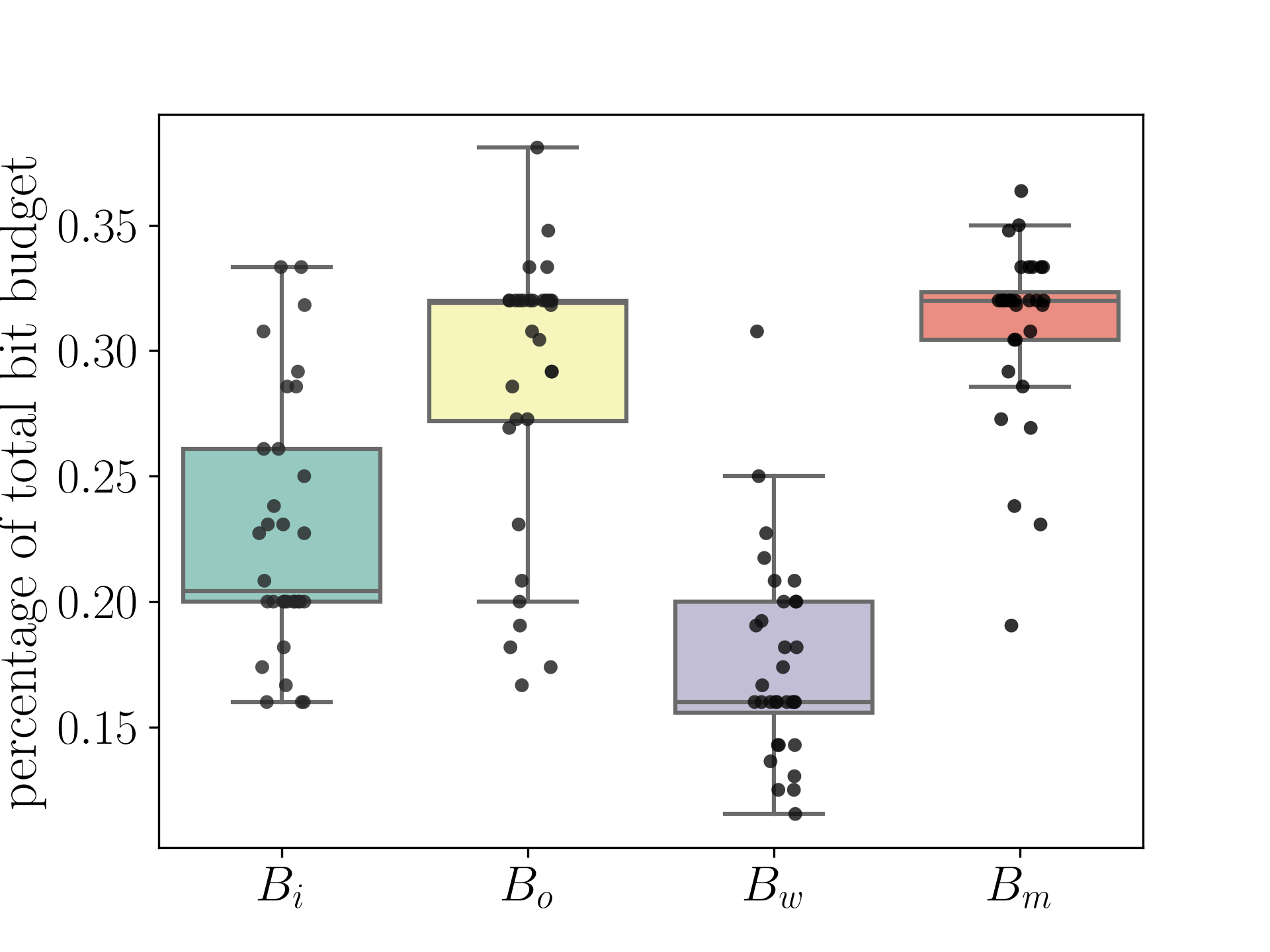}
    \caption{Distributions of bit width controlling the input quantization ($B_i$), output quantization ($B_o$), weights of convolutions ($B_w$) and intermediary activations ($B_m$), in the best-performing models for all approaches, datasets and tasks.}
    \label{fig:bit_parameters}
\end{figure}

We displayed in Figure \ref{fig:bit_parameters} the distribution of bit width parameters in our best-performing quantized models. On average, convolution weights used significantly lower bit widths, representing only about 15\% of the total budget. In contrast, bit widths controlling the output quantization represented more than $35\%$ of the total budget on average. This suggests a heuristic for optimal quantization parameters: choosing higher bit widths on activations and lower bit widths on weights leads to better performances. This aligns with our observation that the poorman's transform, an extreme quantization of STFT coefficients, outperformed the baseline, and other approximations.

\subsection{Limitations}

The quantization needed to compile models to FHE entails the loss of harmonics: as seen in Figure \ref{fig:SpectrogramVocalset}, the energy of frequency bins above 4000 Hz is quantized to 0 for the STFT, Mel filterbanks and gammatone filters. 
Another limitation of this work is that we did not implement very deep models that tackle  speech processing tasks such as speech recognition or speaker diarization. Additionally, it would be insightful to extend our study to more extensive and complex datasets containing noisy speech and acoustic events. 

Finally, although it is not a limitation of our method per se, the current main drawback of FHE is its inference speed. Computing the STFT over a 64 ms long audio in FHE took 12970 seconds on a Apple M2 chip with 8 cores, but only 0.004 seconds in the clear on the same machine.

\section{Conclusion}

In this work, we provided the first secure raw audio processing pipeline using fully homomorphic encryption. We successively demonstrated the validity of our approach across increasingly difficult tasks: first the computation of four standard time-frequency representations, then the computation of four audio descriptors over different representations in one single model, and finally the training of end-to-end privacy-preserving CNN classifiers for both binary and multi-class classification. We further improved the performance of FHE audio descriptors and FHE classifiers by introducing approximations of time-frequency representations which optimize the quantization of our models. Research on improving fully homomorphic encryption algorithms toward production is blooming with projects such as Google's HEIR \citet{heir}, sharing compilers to facilitate FHE compilation at scale.

\bibliographystyle{plainnat}
\bibliography{bibliography}

\appendix

\section{STFT approximations}
\subsection{Calculations for dilation}\label{demo:dilation}

\begin{align*}
X^{(d)}(m, k) &= \sum_{l, ld < N}x[ld]w[ld-mh] e^{-2i\pi k \frac {ld}{N}} \\
 &= \sum_{n}x(n)w(n-mh) e^{-2i\pi k \frac {n}{N}} 1_{n \equiv 0[d]} \\
  &= \sum_{n}x(n)w(n-mh) e^{-2i\pi k \frac {n}{N}} \frac{1}{d}\sum_{j=0}^{d-1}e^{2i\pi n\frac{j}{d}}  \\
&= \frac{1}{d}\sum_{j=0}^{d-1}\sum_{n=0}^{N-1}x(n)w(n-mh) e^{-2i\pi \frac{n}{N}\left( k-\frac {jN}{d}\right)} \\
&= \frac{1}{d}\sum_{j=0}^{d-1}X\left[m, k- \frac {jN}{d}\right]
\end{align*}

\subsection{Calculations for poorman's transform}\label{demo:poorman}
We have

\begin{align*}
    | X(m, k) &- X^L(m, k) | = 
    \left| \sum_{n = 0}^{M-1} x(n) w(n-mh) \left( e^{- \frac{2j\pi kn}{N}} - p_L\left( e^{- \frac{2j\pi kn}{N}} \right) \right) \right|.
\end{align*}
Let $\theta \in [0, 2\pi]$ such that $p_L(e^{i\theta}) = e^{2i\pi  \frac{l}{L}}$.
Then,
\begin{align*}
   \left| e^{i \theta} - p_L(e^{i\theta}) \right| &=  \left| e^{i \left(\frac{\theta}{2} - \pi\frac{l}{L} \right) } - e^{-i \left(\frac{\theta}{2} - \pi\frac{l}{L} \right) } \right| \\
   & = 2\left| \sin \left(\frac{\theta}{2} - \pi\frac{l}{L} \right) \right|.
\end{align*}
By definition of $p_L$, $\left|\theta - 2\pi\frac{l}{L} \right| \leq \frac{\pi}{L}$, then
\begin{equation*}
        \left| e^{i \theta} - p_L(e^{i\theta}) \right| \leq 2\left|\sin \left(\frac{\pi}{2L} \right) \right|.
\end{equation*}
Finally, using the Cauchy-Schwarz theorem,
\begin{equation*}
    \left|X(m, k) - X^{L}(m, k)\right| \leq 2\left|\sin\frac{\pi}{2L}\right| \sqrt{\sum_{i=0}^{M-1}(x(i)w(i-mh))^2}.
\end{equation*}

\end{document}